\documentclass[letters,usenatbib]{mnras}
\usepackage{orcidlink}
\usepackage{newtxtext,newtxmath}
\usepackage[T1]{fontenc}
\usepackage{graphicx}	
\usepackage{amsmath}	
\newcommand{\dl}{\lambda}
\newcommand{\logl}{log~(L/$L_{\odot}$) }
\usepackage{gensymb}

\title[HD\,144812: a post-RSG in a binary system]{HD\,144812: A transition-phase massive star in a binary system}

\author[M. Kourniotis et al.]{
Michalis Kourniotis$^{1}$\thanks{E-mail: kourniotis@asu.cas.cz}\orcidlink{0000-0003-0206-8802},
Michaela Kraus$^{1}$\orcidlink{0000-0002-4502-6330},
Maria Laura Arias$^{2,3}$\orcidlink{0000-0002-4016-2501},
and Lydia S. Cidale$^{2,3}$\orcidlink{0000-0003-2160-7146}
\\
$^{1}$Astronomical Institute, Czech Academy of Sciences, Fri\v{c}ova 298, 251\,65 Ond\v{r}ejov, Czech Republic\\
$^{2}$Instituto de Astrof\'{\i}sica La Plata, CCT La Plata, CONICET-UNLP, Paseo del Bosque S/N, B1900FWA, La Plata, Argentina\\
$^{3}$Departamento de Espectroscop\'{\i}a, Facultad de Ciencias Astron\'omicas y Geof\'{\i}sicas, Universidad Nacional de La Plata (UNLP),\\ Paseo del Bosque S/N, B1900FWA, La Plata, Argentina
}

\date{Accepted 2025 March 24. Received 2025 March 20; in original form 2025 January 10}

\pubyear{2025}

\begin{document}
\label{firstpage}
\pagerange{\pageref{firstpage}--\pageref{lastpage}}
\maketitle

\begin{abstract}
In this Letter, we shed light on the evolutionary phase of HD\,144812, a Galactic yellow supergiant showing infrared excess that is typically expected for evolved stars undergoing enhanced mass-loss activity. We present high-resolution spectroscopy of the star in the $H-$ and $K-$band acquired with the GRating INfrared Spectrometer (IGRINS) and further explore multi-band imaging of the wider field of view from the ultraviolet to the radio regime. The IGRINS data reveal several lines from the hydrogen series and iron in a double-peaked emission and we here suggest, that HD\,144812 is orbited by a disk-hosting companion. Furthermore, we report emission in the CO band heads of the star that is modeled to arise from a circum-stellar/binary disk (or ring) of ejected gas. The latter consists of material that is expected to have been dredged up from the core of the star to its surface during a prior phase as a red supergiant (RSG). These findings together suggest that HD\,144812 is a rare, post-RSG star in a binary system, encouraging further investigation on the effect that the stellar encounters have on triggering instabilities and driving the evolution of the primary star shortly prior to the supernova event.
\end{abstract}


\section{Introduction}

Modeling the late stages of massive stars is an intriguing subject in
stellar astronomy. The scarcity of observational data, however, makes it challenging to validate the various evolutionary scenarios. In addition, the physics of the enhanced, often episodic, mass loss that is tightly associated  with the advanced stellar stages is still not well understood. The need for explaining instability in connection with mass eruptions prior to the supernova (SN) event is imperative, as it plays a decisive role in altering the appearance of the stars and defining their ultimate fate. The released mass and energy resulting from such events further modulates the chemistry and structure of the interstellar medium.  

The established concept that a large fraction of massive stars interact with companion stars \citep{2013A&A...550A.107S} adds on the complexity of the evolutionary puzzle. Through mass transferring and envelope stripping, properties of the binary components such as the rotation, surface abundances, and emitting radiation can be drastically altered \citep{2012ARA&A..50..107L,2013ApJ...764..166D,2019A&A...629A.134G}. In the case of merging stars, the product is a hybrid object with a reformed interior and peculiar chemical composition \citep{2024ApJ...967L..39B,2024ApJ...963L..42M} that can evolve and terminate its life against the theoretical expectations of a single-evolving counterpart \citep{1992ApJ...391..246P,2014ApJ...796..121J}. The subject of post-interaction evolution has been studied thoroughly for the early types \citep[e.g.][]{2013ApJ...764..166D,2023A&A...672A..22B}, yet disanalogous progress has been made for stars in the red-supergiant (RSG) stage and beyond. This is primarily due to the significant drop in the intrinsic binary fraction among RSGs \citep[e.g.][]{2022MNRAS.513.5847P}, which is in line with the suggestion that a large fraction  of the early post-main-sequence stars are merger products \citep{2024ApJ...963L..42M}. The instrumental limitations (i.e., angular, spectral and temporal  resolution,  low signal-to-noise ratio, dynamic range of CCD detectors, etc) and the insufficient observing coverage also contribute to the low detection rate of evolved star multiples, particularly those that underwent significant widening of their orbit \citep{2005A&A...435.1013P}. Yet, capturing massive stars in binaries shortly prior to the core collapse is of paramount importance for constraining the modeling of the pre-SN channels \citep[e.g.][]{2021A&A...645A...5S}. Importantly, when the parameters of such systems raise the possibility of post-SN merging between the remnants, these objects manifest themselves as candidate progenitors of gravitational-wave sources \citep{2014LRR....17....3P}.

The evolution of massive stars beyond the main sequence is rapid, which prevents us from observing the stars in transition, exotic phases. On top of their scarcity per se, the transition-phase massive stars are often misclassified as steadily-evolving (``normal'') stars on the basis of limited known parameters. A characteristic such example is that of yellow hypergiants (YHGs), namely luminous AFG-type stars that exhibit evolution \textit{bluewards} as post-RSGs. Given that they occupy the same  region on the evolutionary diagram with the most luminous (\logl$\gtrsim5.4$) yellow supergiants (YSGs) i.e. core-helium burning objects moving coolwards for the first time, a classification based solely on the spectral type is ambiguous. Another transition type, the luminous blue variables (LBVs), may also emerge in the YSG regime when they possess optically-thick outflows\footnote{We refer to the so called ``low-luminous'' LBVs with \logl$\lesssim5.8$, which are believed to be post-RSG stars \citep{2016ApJ...825...64H}.}, following phases of episodic mass loss \citep[see Fig. 6 in][]{2015MNRAS.447..598S}. In principle, the solid identification of a post-RSG requires to assess signatures of an evolved and transient status, which includes a peculiar chemical profile, distinctive (e.g. strange-mode) oscillations, and essentially, the high-amplitude spectrophotometric variability caused by instability of the stellar atmosphere \citep{2024MNRAS.529.4947G}. A classification as a post-RSG star can be augmented by inspection of the surrounding field, in search of traces of the outer envelope shed during a prior RSG phase and of signs of interaction between a highly energetic stellar feedback and the interstellar medium.

As part of an ongoing effort to increase the number of known post-RSGs in the Milky Way and nearby galaxies, we have been carrying out spectroscopic observations of YSGs from Simbad with infrared (IR) excess. The latter feature could emerge potentially from a dusty circumstellar envelope being either the relic from prior shedding of gas during the RSG phase or, that was recently formed following intensive mass-loss episode(s). In this Letter, we discuss the case of HD\,144812 (R.A.\,16$^h$09$^m$59.15$^s$, Dec.\,-48$\degree$34$\arcmin$27.94$\arcsec$; J2000), a Galactic emission-line YSG \citep[F3Iae;][]{1993yCat.3051....0H} with IR excess ($K_{s}-W4=1.9$ mag), which was selected for high-res spectroscopic investigation. Here, we identify HD\,144812 as a post-RSG star in a binary system, placing the object among the few documented stars in transition phase that are also observed as stellar multiples. The study is based on $HK-$band spectroscopy and archival multi-band imaging, which are described in Sect. \ref{sec:data}. In Sect. \ref{sec:res}, we update the spectral classification of HD\,144812, discuss the emission features that we propose to emerge from a companion disk, proceed with the modeling of the CO emission, and highlight the prominent features of the surrounding field. The findings are jointly discussed in Sect. \ref{sec:disc}.

\section{Observations \& archival data}
\label{sec:data}

We obtained near-IR spectroscopy of HD\,144812 on March 9, 2024 (JD 2460378.7168) with the Immersion GRating INfrared Spectrometer \citep[IGRINS;][]{2014SPIE.9147E..1DP} mounted on the 8.1-m Gemini South telescope, Chile, under the Gemini Program ID GS-2024A-Q-307. The instrument delivers high-resolution (R$\sim$45\,000) echelle spectra over the $H$- and $K$-band ($1.45-2.45$ $\mu$m). The observations were taken with an ABBA pattern and the data reduction (background subtraction, flatfield, aperture extraction, wavelength calibration) was performed using the dedicated IGRINS pipeline. The mean signal-to-noise of the reduced data is 150. A telluric standard star of A0V spectral type was observed close in time and at similar airmass to the target exposure; the telluric correction of the science spectra was done using the relevant task from \texttt{IRAF} package\footnote{\texttt{IRAF} was distributed by the National Optical Astron-
omy Observatory, which was managed by the Association of
Universities for Research in Astronomy (AURA) under a co-
operative agreement with the National Science Foundation.}. We performed continuum normalization by fitting the data with low-order polynomials. Finally, the spectra were corrected to the heliocentric reference frame by applying to these a velocity shift of 26.7 km~s$^{-1}$. 
 
We explored the wider field ($20\arcmin\times20\arcmin$) around HD\,144812 searching for signs of interaction of the star with the interstellar medium. We made use of the \textsc{skyview} package, which is included in the \textsc{python} module \textsc{astroquery} \citep{2019AJ....157...98G}, for querying data produced by various imaging surveys. 
Images of the target field were taken in the near-UV (Galaxy Evolution Explorer, GALEX),  optical (red) range (Digitized Sky Survey, DSS2), near-IR (Two Micron All-Sky Survey, 2MASS J, H, K$_{s}$ bands), mid-IR (Wide-field Infrared Survey Explorer, WISE W1, W2, W3, W4 bands), far-IR (AKARI satellite, 60 and 160 $\mu$m bands) and in the radio (Sydney University Molonglo Sky Survey, SUMSS at 843 MHz). Other queried data in the optical include images from the Transiting Exoplanet Survey (TESS; obtained with a wide bandpass centered at 787 nm) and from the Southern H-Alpha Sky Survey Atlas (SHASSA; exploring both the H$_{\rm \alpha}$ and the continuum-subtracted/corrected images). 

HD\,144812 was also observed by the Radial Velocity Experiment (RAVE) spectroscopic survey \citep{2020AJ....160...82S}, which provided radial velocity measurements of southern sources from the study of the \ion{Ca}{} triplet ($8410-8795$ \AA). Four observations of the object were conducted on $07-10$ May 2004 ($-48\pm9$ to $-39\pm9$ km~s$^{-1}$, respectively) and three, later on that month, on 30 May 2004 ($-7\pm4$ km~s$^{-1}$). The RAVE spectra display strong \ion{Ca}{II} absorption lines (comparing to the Paschen ones) that typically identify late-type objects. Interestingly, an earlier, single measurement for HD\,144812 was obtained on 6 August 2003. When comparing to the data of 2004, the radial velocity of the star in 2003 was significantly offset ($+76\pm18$ km~s$^{-1}$), with the data showing weaker \ion{Ca}{} lines that suggest a higher effective temperature. The latter spectrum of HD\,144812 was flagged by the survey for possible effects of binarity.

\section{Results}
\label{sec:res}

In Fig. \ref{fig:spectra}, we present the IGRINS $H$- (upper panel) and $K$-band (lower panel) spectra of HD\,144812. Both are characterized by a wealth of photospheric absorption lines, containing also distinctive emission features mostly from the hydrogen series and from iron. The high-res $K-$band data also enable to resolve the rotationally broadened band heads of the first-overtone CO band emission and individual ro-vibrational lines.

\subsection{Spectral classification}

Absorption lines of neutral metals in the IGRINS spectra of HD\,144812 suggest a late spectral type, which is in line with the documented F-type classification of the star. For estimating the effective temperature, we compared the observations with synthetic data of warm supergiants in the infrared. We computed the theoretical spectra using the radiative transfer and spectral synthesis code Turbospectrum via the framework for spectroscopic data analysis iSpec \citep{2014A&A...569A.111B}. The calculations by Turbospectrum were performed with 1D MARCS model atmospheres, under the assumption of local thermodynamic equilibrium (LTE). The metallicity of the models was chosen to be solar, with the abundances been taken from \cite{2007SSRv..130..105G}. The input line list in the infrared was extracted from the Vienna Atomic Line Database\footnote{https://vald.astro.uu.se/}. 

We computed model spectra for $T_{\rm eff}=6\,000-7\,000$ K with a step of 100 K, at the lowest grid value for the surface gravity, log\,$g=1.0$ (or 1.5, for $T_{\rm eff}=7\,000$ K). The resolution of the spectra was set to that of the IGRINS data. The theoretical lines were subject to a micro- and macro-turbulence broadening of 10 and 15 km~s$^{-1}$, respectively, being typical values for high-luminous AFG-type stars with extended atmospheres \citep{2014ARep...58..101K,2022MNRAS.511.4360K}. We considered that line broadening due to the stellar rotation is negligible. In Fig. \ref{fig:spectra}, an emission-free portion of the observations is enclosed by a dashed line, and is shown in the inset figures on the lower left of the two panels. In the inset figures, the IGRINS data (black) are overplotted by the best-fit MARCS model of 6\,400 K (cyan), which characterizes the star as a mid F-type \citep{2010MNRAS.402.1369L}. For comparison, the model spectra of 6\,000 K (green) and 6\,800 K (blue) are also displayed. When compared to the fit model, the science spectra were found to be blueshifted by $5\pm1$ km~s$^{-1}$.

\subsection{Emission lines}
\label{emission}

We report emission for various spectral lines\footnote{Following our data processing, the rest wavelengths are given in vacuum.}, several of which are resolved with a double-peaked profile. The latter features include three \ion{H}{I} lines from the Brackett series $\dl\dl1.6811\,(4-11),1.7367\,(4-10)$, and the Br$\gamma$ $\dl2.1661\,(4-7)$, as well as lines of \ion{Fe}{II} $\dl\dl1.6878, 1.7420, 2.0892$, and [\ion{Fe}{II}] $\dl1.6440$ (possibly also at $2.0466\,\mu$m). The data also show a prominent and highly-asymmetric emission at 2.0587 $\mu$m from the \ion{He}{I} transition; a two-component feature is likely to emerge on top of a broad line core, with a peak-to-peak separation of approx.~130 km~s$^{-1}$.  We display these lines on the right panel of Fig. \ref{fig:spectra}. Broad emission is also detected for various metallic lines, which were fitted with a Gaussian function\footnote{We excluded from the process the blended $\dl1.5044$ and the weak $\dl2.1438$.} to determine the full width at half maximum (FWHM). These lines are \ion{Mg}{I} $\dl\dl1.5044,1.7113,2.1067$ (FWHM = $320\pm30$ km~s$^{-1}$), \ion{Mg}{II} $\dl\dl2.1375,2.1438$ (FWHM = 285 km~s$^{-1}$), and the \ion{Na}{I} doublet, $\dl\dl2.2062, 2.2090$ (FWHM = 260$\pm$15 km~s$^{-1}$), which further display superimposed photospheric (narrow) absorption at the relevant velocity frame. Finally, we attribute emission at $\sim$2.190 $\mu$m to \ion{He}{II} $(10-7)$ with a FWHM of 290 km~s$^{-1}$, although the latter value could be misjudged due to the presence of adjacent photospheric absorption and to possible contribution from \ion{Fe}{I} $\dl2.1901$. Weak and broad emission possibly from \ion{Fe}{I} is also seen at $\sim$1.5297 and $\sim$2.1783 $\mu$m.

The depression in the core of the Br11 \& Br10 lines extending below the continuum level (Fig. \ref{fig:spectra}, right) is characteristic of a shell-type line profile, which is seen in stars surrounded by highly-inclined (edge-on) gaseous disks \citep{2003PASP..115.1153P}. Double-peaked emission with a strong line core is also observed for the Br$\gamma$ line, with the violet component being stronger than the red, sign of asymmetries within the disk or of an additional blue-shifted emission component. The differences in the \ion{H}{I} profile between the two spectral bands can be attributed to the different emission regions of these lines, with the Br11, having a higher upper energy level (13.486 eV), arising from the inner disk radii \citep{2015AJ....149....7C}. Under the assumption of the disk being Keplerian, this effect can be also seen from the differences in the peak-to-peak velocity separation; the latter is measured for the Br$\gamma$, Br10 and Br11 lines as 80, 95, and 100 km~s$^{-1}$, respectively. We further report the centroid of the hydrogen emission to be modestly blueshifted from the rest wavelength (i.e.\,the photospheric frame of HD\,144812) by 3$\pm$1 km~s$^{-1}$. 

A double-peaked emission from \ion{Fe}{II} is among the most frequently observed features in the spectra of Be stars \citep[e.g.][]{2015AJ....149....7C,2000A&AS..141...65C,2006A&A...460..821A}. The line is excited by Ly$\alpha$ fluorescence in a dense and warm ($T\sim5000$ K) environment. We measured a peak-to-peak separation of 44$\pm$4 km~s$^{-1}$, which suggests that the line emerges outside the \ion{H}{I}-emitting region, keeping also in mind that this value can be even lower due to interference of the feature with the photospheric \ion{Fe}{II} absorption of HD\,144812. The optically thin [\ion{Fe}{II}] emission line also displays a velocity separation of 45 km~s$^{-1}$. With the latter line arising from the extended shell (essentially tracing gas with low density), this agreement with the \ion{Fe}{II} kinematics suggests a possible deviation from the Keplerian motion in the outer disk. With the exception of \ion{Fe}{II} $\dl2.089$, the iron emission core appears to be blueshifted from the \ion{H}{I} line core. This effect, which has been also reported in \cite{2015AJ....149....7C} for Be stars, can be explained by the existence of one-armed density waves in the disk \citep{1991PASJ...43...75O} causing displacements of the metallic emission in the direction of the strongest of the two \ion{H}{I} peaks, here being the violet.

The situation is less clear with the remaining emission features. None of these, broad lines, exhibit a disk-like line profile implying that, if they arise in the disk environment, they should be either optically thick or that their formation region might spread over a velocity range that smears out the specific double-peaked shape. On the other hand, fluorescent emission from low-excitation lines such as of \ion{Na}{I} requires that the gas is shielded against the radiation of the central source \citep[in order to prevent the ionization of the metal;][]{1988ApJ...324.1071M}, hence the line is expected to trace the outer radii. Yet, the large kinematical broadening of \ion{Na}{I}, which exceeds considerably that of \ion{H}{I}, renders the site of excitation debatable.

\subsection{Modeling of the CO bands}

\begin{figure*}
   \centering
   \includegraphics[width=18cm]{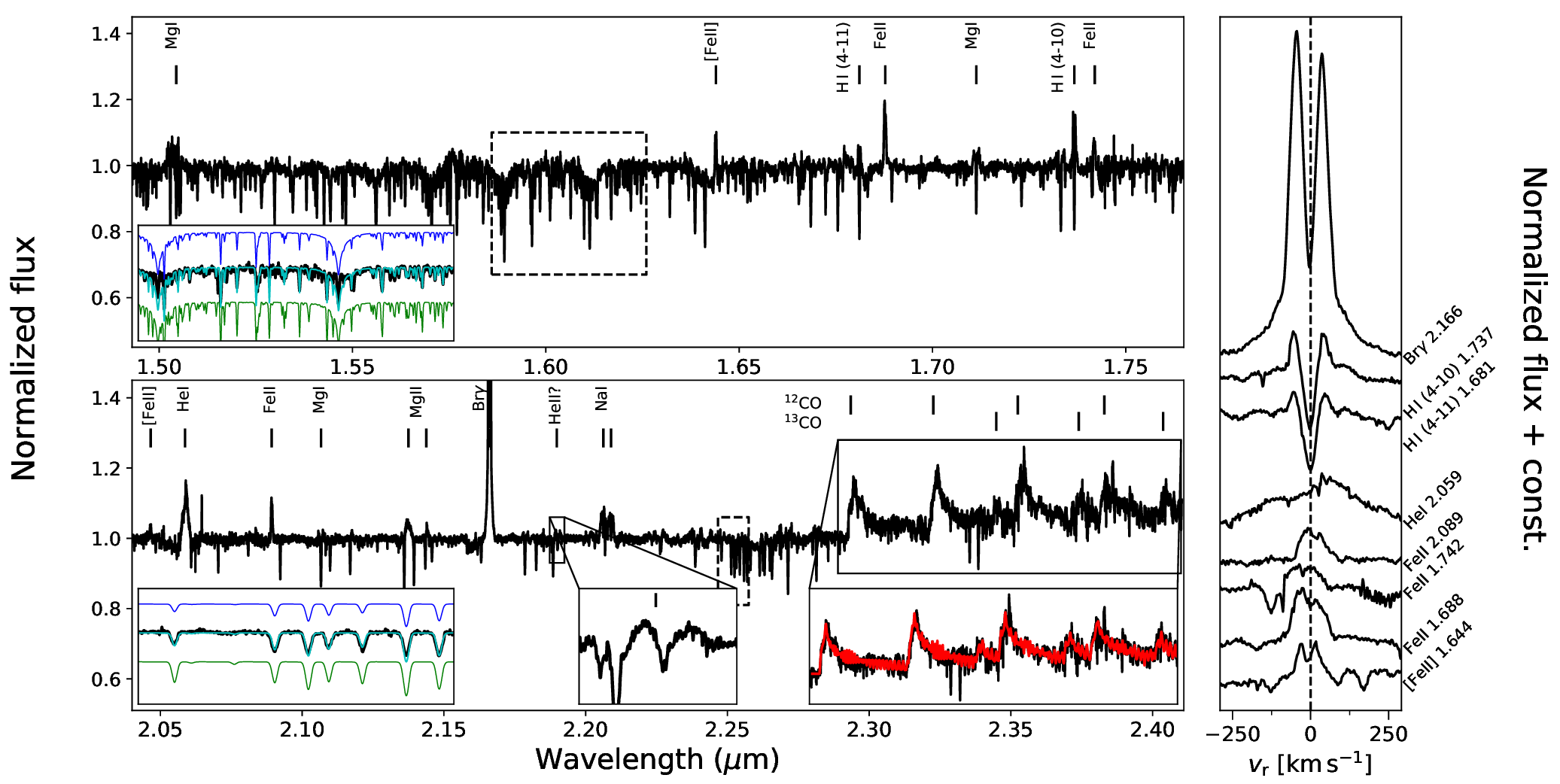}
      \caption{\textit{Left.} IGRINS $H$- (upper panel) and $K$-band (lower panel) spectra of HD\,144812 (black). Selected regions of the spectra (dashed-line boxes) are displayed in the left inset figures. Within the latter, we display three MARCS model spectra of 6\,000 K (green), 6\,400 K (cyan - best fit) and 6\,800 K (blue). In the central inset of the lower panel, we display a zoomed view of the \ion{He}{II} $\dl2.1898$ line. In the right inset, we show the spectral region containing the CO bands, with overplotted the best-fit molecular disk/ring model (red). \textit{Right.} Identified double-peaked emission lines, which are displayed in velocity scale around the rest wavelength (vertical dashed line).}
         \label{fig:spectra}
\end{figure*}

The most pronounced feature in the $K-$band spectrum of HD\,144812 is the emission from the first-overtone bands of the CO molecule, with four band heads of the $^{12}$CO isotope and three of $^{13}$CO (Fig. \ref{fig:spectra}). The  observed lines display a blueshift by $3\pm2$ km~s$^{-1}$ from the  photospheric frame of the luminous star. Collisional excitation gives rise to the CO emission that traces dense, cool gas ($T_{\rm CO}\leq5000$ K). The broad band heads display a blue shoulder and a red peak, characteristic of rotation, being suggestive of a disk/ring structure that enables the molecular gas to retain low temperatures within an intense radiation field. We modeled the profiles of the individual ro-vibrational lines by a body with such a geometry and under the assumption of LTE, using the molecular code by \cite{2000A&A...362..158K} and including calculations for the $^{13}$CO isotope \citep{2009A&A...494..253K,2013A&A...558A..17O}. Line broadening was assumed to be predominantly due to disk/ring rotation. For modeling the turbulent motion of the gas, a Gaussian line profile was used.    

We list the parameters that best fit the IGRINS data in Table \ref{tab:COpar}, and display the corresponding model (in red) in Fig. \ref{fig:spectra} (lower panel, right inset). Highlighted is the enrichment of the ejected gas in $^{13}$CO, as this is inferred from the low carbon isotope ratio $^{12}$C/$^{13}$C = $2.5\pm0.5$. In turn, the high abundance of $^{13}$C in the surface, being dredged-up from the core when the star was in the RSG phase, traces an advanced evolutionary stage  \citep[e.g.][]{2010MNRAS.408L...6L,2013A&A...558A..17O}.

\begin{table}
\centering
\caption{\label{tab:COpar} Best-fit parameters from the CO emission modeling of HD\,144812.} 
\begin{tabular}{l|r}
\hline
Parameter & Value \\
\hline\hline
$T_{\rm CO}$ & $3500\pm200$ K \\
$N_{\rm CO}$ & $8\pm2\times10^{20}$ cm$^{-2}$ \\
$^{12}$C/$^{13}$C & $2.5\pm0.5$ \\
$\varv_{\rm rot, proj}$ & $120\pm10$ km~s$^{-1}$ \\
$\varv_{\rm gauss}$ & $20\pm5$ km~s$^{-1}$ \\
\hline
\end{tabular}
\end{table}

\subsection{The environment of HD\,144812}
\label{ssec:env}

\begin{figure*}
   \centering
   \includegraphics[width=14.9cm]{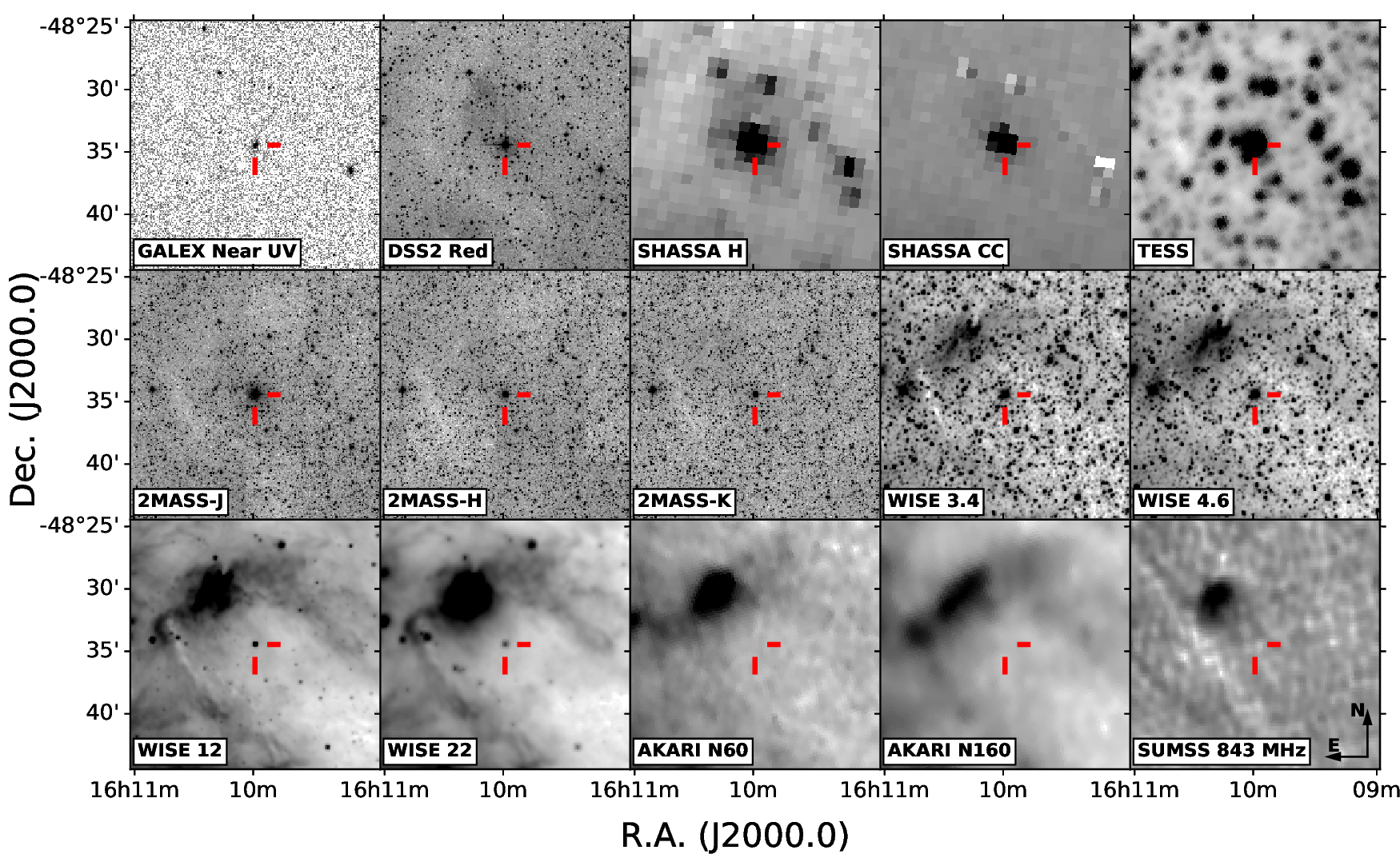}
      \caption{Multi-band imaging ($20\arcmin\times20\arcmin$) of the field around HD\,144812. From top to bottom, left to right, we show data from the surveys GALEX, DSS2, SHASSA, TESS, 2MASS, WISE, AKARI, and SUMSS. The red bars point to the target and have a size of 1$\arcmin$. North is up, east is left.}
         \label{fig:images}
\end{figure*}

In Fig. \ref{fig:images}, we present the $20\arcmin\times20\arcmin$ images of the field around HD\,144812 in different bands. The target stands out as a prominent source in the near-UV, optical, near- and mid-IR field up to 22 $\mu$m. The latter data suggests that the luminous star is embedded within a dust envelope. Imaging from WISE additionally reveals a pronounced feature NE of the target, which is also detected as radio continuum emission at 843 MHz, suggesting thermal emission from warm dust \citep{2012A&A...542A..10A}. Emission from the wider field at the WISE longer wavebands appears to surround a low-density area (or cavity) that hosts the studied object. 
 
The continuum-corrected H$\alpha$ imaging (Fig. \ref{fig:images}) traces recombination emission from the close environment of the star. Diffuse emission, mostly in the N/NE direction, extends $5.5\arcmin$ from the target. The feature can be also seen in the DSS2 image. Adopting the \textit{Gaia} distance\footnote{We note that the \textit{Gaia} renormalised unit weight error of the star is 1.4, indicating marginally reliable astrometry e.g. due to  unresolved binarity of the target. Hence the distance value should be taken with caution.} to the star of 1336 pc \citep{2021yCat.1352....0B}, the H${\rm \alpha}$-emitting area has a projected size roughly --due to the low resolution-- of 2 pc. 

\section{Discussion and conclusions}
\label{sec:disc}

HD\,144812 displays a mid-F spectrum with superimposed double-peaked emission lines (Fig. \ref{fig:spectra}), which are commonly seen in the data of disk-hosting objects. The possible contribution to the disk spectrum from \ion{He}{I} $\dl2.0587$ predisposes then to the presence of a hot companion, with a spectral type earlier than B2.5 \citep{2000A&AS..141...65C}. Moreover, a hot star companion can also explain the detection of the object on the UV image (Fig. \ref{fig:images}). However, direct evidence in the spectra that would pinpoint the case of a classical Be star possessing a rotationally-induced outflow are missing (or masked by the flux of the primary). Instead, we consider as credible the scenario that the double-peaked features arise from an \textit{accretion} disk, which consists of material that is channeled from HD\,144812 to the companion via a (recent or ongoing) Roche lobe overflow. In support of this case, the broad, optically thick \ion{He}{II} $\dl2.1898$  (Fig. \ref{fig:spectra}; lower panel, central inset) suggests its origin in a high-excitation region that can be powered by the accretion flow, as seen also in various X-ray binaries \citep[e.g.][]{1999A&A...348..888C,2012A&A...544A.149M}. 

Modeling the CO emission favors the scenario that HD\,144812 is found in a post-RSG phase evolving bluewards. Processed, ejected gas from the surface of the star emits in the form of disk (or ring) and it needs to be determined whether the latter body corresponds to the companion disk, whether it traces circumstellar material of the super(hyper)giant, or it is circumbinary in nature. Supportive to the first case would be the observed blueshift of the CO emission lines from the velocity frame of HD\,144812, which is similar to that of the hydrogen ones. In this scenario, however, the molecular gas has to be found in the outer radii, as the radiation field of the hot disk is expected to be repressive for the CO formation. On the other hand and similar to the \ion{Na}{I} line, the large kinematical broadening of CO (Table \ref{tab:COpar}) exceeding that of \ion{H}{I}, would place the molecule in the inner radii, unless there is evidence of enhanced transport of angular momentum into the outer shell. We note here that, the observed displacements in the emission features are regarded as modest, and more observations are needed in order to disentangle the distinct kinematics. The same conclusion applies also for the scenario that CO spans a circumbinary disk, which can be formed potentially from gas that escapes the system through the Lagrangian point L$_{2}$ \citep{2018MNRAS.479.4844C}.  

The scenario that CO emission originates in a shell around HD\,144812 is plausible. The molecule is frequently detected in the environment of evolved stars, especially B[e] supergiants (B[e]SGs) and YHGs, serving as a robust indicator of their evolutionary stage \citep[e.g.][]{2018MNRAS.480.3706K,2019Galax...7...83K,2023Galax..11...76K}. Moreover, emission from \ion{He}{I}, \ion{Mg}{II}, and \ion{Na}{I} is found in the spectra of B[e]SGs, as well as of LBVs such as AG Car, HR Car, and HD\,316285, where they arise in their ionized winds and extended envelopes structured with prior-ejected material \citep{1988ApJ...324.1071M,1996ApJ...470..597M}. The detection of \ion{Mg}{I} is not a common feature for the hotter objects in transition phases. The line rather traces a neutral \ion{H}{I} region and thus, it is detected in spectra such as of, the YHG IRC+10420 \citep{2002AJ....124.1026H}, its analogue IRAS 1835-06 \citep{2014A&A...561A..15C}, and the cool LBV W243 \citep[A3Ia$^{+}$;][]{2009A&A...507.1597R}. We note here that, none of the above types show evidence of \ion{He}{II} due to their unfavorable low-temperature conditions, which strengthens our hypothesis that the detection of the latter line in the data of HD\,144812 originates likely in an accreting component.

The possibility that HD\,144812 undergoes energetic mass-loss phases, encouraging also the formation/maintenance of a CO disk from the ejected material, is raised by inspecting the wider field of the target (Fig. \ref{fig:images}). Weak mid-IR emission within, at least, 3$\arcmin$ from the target compared to the outer emission field, could point to dust clearance/sputtering by propagating shock wave(s); evidence of material being shocked is seen in the N/NE direction. A field structured with filaments in the SE/NW (yet not in SW) would be then explained by the ejecta being asymmetric \citep[e.g.][]{2001AJ....121.1111S,2009A&A...507..301D,2018MNRAS.477L..55A}. In the same vein, the warm dust blob at NE of HD\,144812 is likely to trace post-shock gas that is swept up by the stellar feedback. 

The yellow super(hyper)giant HD\,144812 complements the sparse sample of post-RSGs identified in binary systems, offering valuable insights into the late evolutionary stages of massive stars. In the AFG spectral group where the object resides, the only thoroughly studied such case is the YHG HR5171 A \citep{2014A&A...563A..71C}. Near-IR interferometry along with time-series photometry of the star revealed the presence of a low-mass companion that is engulfed within the dense wind of the hypergiant. Their interaction was suggested by the authors to be decisive for the evolution and  fate of the evolved primary. More recently, \cite{2022MNRAS.511.4360K} reported the discovery of a disk-hosting companion orbiting the YHG HD\,269953 in the Large Magellanic Cloud (LMC). Using multi-epoch optical spectroscopy, the star was found to display double-peaked \ion{Fe}{I} emission with Doppler shifts due to the orbital motion of the companion. Upper atmospheric emission from the hypergiant was shown to be coupled to the disk kinematics, implying a possible interplay between the circumstellar environment of HD\,269953 and the companion. It is noteworthy that, the optically thick H$\alpha$ in the spectrum of HD\,269953 showed no evidence of the disk and captured, instead, the infall of cooling gas from the upper atmosphere of the hypergiant. As with the case of HD\,144812, the presence of a CO molecular disk around the LMC star was later confirmed using near-IR spectroscopy \citep{2022BAAA...63...65K}.

We aim to conduct monitoring of HD\,144812 in both the infrared and optical in order to disentangle the kinematics and refine the properties of the components. Long-term observations are also required to record unstable mass loss and changes in the opacity of the circumstellar envelope, that would help to conclude whether the star is a YHG or an LBV that is captured in outburst. With the common view that the post-RSG instability is attributed to the acute drop in the stellar mass during the RSG phase or to a distinct core-envelope structure (e.g.~a strongly diluted envelope), evidence of the assistive or inductive influence by unseen companions will add a vital constraint to the evolutionary modeling of these objects.  

\section*{Acknowledgements}

We thank the anonymous referee for providing constructive comments that improved the quality of this work. This paper is based on observations obtained at the international Gemini Observatory, a program of NSF NOIRLab, which is managed by the Association of Universities for Research in Astronomy (AURA) under a cooperative agreement with the U.S. National Science Foundation on behalf of the Gemini Observatory partnership: the U.S. National Science Foundation (United States), National Research Council (Canada), Agencia Nacional de Investigaci\'{o}n y Desarrollo (Chile), Ministerio de Ciencia, Tecnolog\'{i}a e Innovaci\'{o}n (Argentina), Minist\'{e}rio da Ci\^{e}ncia, Tecnologia e Inova\c{c}\~{o}es (Brazil), and Korea Astronomy and Space Science Institute (Republic of Korea), under program ID GS-2024A-Q-307.

This project has received funding from the European Union's Framework Programme for Research and Innovation Horizon 2020 (2014-2020) under the Marie Sk\l{}odowska-Curie Grant Agreement No. 823734. The Astronomical Institute of the Czech Academy of Sciences is supported by the project RVO:67985815. 
MLA and LSC acknowledge financial support from CONICET (PIP 1337) and the Universidad Nacional de La Plata (Programa de Incentivos 11/G162), Argentina.

\section*{Data Availability}
The data underlying this article will be shared on reasonable request to the corresponding author.

\label{lastpage}
\end{document}